\documentstyle[12pt]{article}

\begin{document}


\def\a{\alpha}
\def\b{\beta}
\def\c{\varepsilon}
\def\d{\delta}
\def\e{\epsilon}
\def\f{\phi}
\def\g{\gamma}
\def\h{\theta}
\def\k{\kappa}
\def\l{\lambda}
\def\m{\mu}
\def\n{\nu}
\def\p{\psi}
\def\q{\partial}
\def\r{\rho}
\def\s{\sigma}
\def\t{\tau}
\def\u{\upsilon}
\def\v{\varphi}
\def\w{\omega}
\def\x{\xi}
\def\y{\eta}
\def\z{\zeta}
\def\D{\Delta}
\def\G{\Gamma}
\def\L{\Lambda}
\def\F{\Phi}
\def\P{\Psi}
\def\S{\Sigma}

\def\o{\over}

\def\IJMP{Int.~J.~Mod.~Phys. }
\def\MPL{Mod.~Phys.~Lett. }
\def\NP{Nucl.~Phys. }
\def\PL{Phys.~Lett. }
\def\PR{Phys.~Rev. }
\def\PRL{Phys.~Rev.~Lett. }
\def\PTP{Prog.~Theor.~Phys. }
\def\ZP{Z.~Phys. }

\def\beq{\begin{equation}}
\def\eeq{\end{equation}}


\title{
  \begin{flushright}
    \large UT-794
  \end{flushright}
  \vspace{5ex}
  Primary Inflation}
\author{Izawa K.-I. \\
  \\  {\sl Department of Physics, University of Tokyo} \\
  {\sl Tokyo 113, Japan}}
\date{October, 1997}
\maketitle

\begin{abstract}
We consider an inflationary universe scenario with multiple stages
of inflation.
The primary inflation, which may start at the Planck epoch,
is followed by secondary inflations, which include the cosmological
inflation that causes the primordial density fluctuations of our universe.
We point out that an initial condition for a secondary inflation
is naturally realized if the $e$-fold number of the primary inflation
is sufficiently large.
\end{abstract}

\newpage

The universe, when `created', could have an energy density of order the Planck
scale,%
\footnote{We adopt a unit where the Planck scale is one.}
below which it may be described in terms of classical spacetime.
A chaotic inflation%
\footnote{A chaotic inflation may be naturally realized in supergravity
by means of a hybrid potential with a Fayet-Iliopoulos $D$ term
\cite{Ste}.}
is a plausible candidate which first inflates the `created' universe,
since it may naturally start at the Planck epoch
\cite{Lin}.
However, without fine tuning of coupling constants,
it tends to produce too large density fluctuations
to be identified with the primordial ones observed
in our universe.

Thus we are led to consider
an inflationary universe scenario with multiple stages of inflation
(for example, see Ref.\cite{Iza}).
The primary inflation,%
\footnote{The primary inflation is not necessarily of the chaotic type,
though a chaotic inflation seems adequate.}
which may start at the Planck epoch,
is followed by secondary inflations, which include the cosmological
inflation that causes the primordial density fluctuations of our universe.
Then the primary inflation is free from constraints imposed by
the observed primordial density fluctuations.

When the scale of a secondary inflation is much smaller than the
Planck scale, as is the case for the primordial inflation,
we need a homogeneous region of the horizon size $H^{-1}$ in the universe
which provides a seed for the secondary inflation
(of the new or hybrid type
\cite{Lyt}).
Here $H$ denotes the Hubble parameter during the secondary inflation.

Let $\D \v$ be an allowable range for the initial value of an inflaton
field $\v$ of the secondary inflation.
Then the spacetime derivatives $\q \v$ of the inflaton field in a seed region
for the secondary inflation should be smaller than $H \D \v$.
On the other hand, the chaotic initial condition implies
$\q \v$ of order the Planck scale.

Fortunately, this discrepancy is eliminated by
the primary inflation if the $e$-fold number $N'$ of the primary inflation
satisfies the following condition:
\beq
 e^{-N'} < H \D \v.
\eeq
Hence we conclude that an initial condition for
the secondary inflation may be naturally achieved in a seed region
as a bonus of the primary inflation,%
\footnote{We assume that the sector of
the primary inflation is properly separated
from that of the secondary one.
For example, an $F$-term inflation in supergravity affects
initial values of inflatons in the subsequent inflations
\cite{Iza},
which is beyond the scope of the present arguments.}
which first inflates the `created' universe at the Planck epoch.

The above observations imply that
the picture of multiple inflations
may be relevant for understanding the present status of our universe.

\section*{Acknowledgements}

The author would like to thank T.~Yanagida for valuable discussions.


\newpage

\begin{thebibliography}{99}

\bibitem{Ste}
  E.D.~Stewart, \PR {\bf D51} (1995) 6847; \\
  P.~Bin{\' e}truy and G.~Dvali, \PL {\bf B388} (1996) 241; \\
  E.~Halyo, \PL {\bf B387} (1996) 43; \\
  See also D.H.~Lyth and A.~Riotto, hep-ph/9707273; \\
  D.H.~Lyth, hep-ph/9710347.

\bibitem{Lin}
  A.D.~Linde, {\sl Particle Physics and Inflationary Cosmology}
  (Harwood Academic Publishers, 1990).

\bibitem{Iza}
  Izawa~K.-I., M.~Kawasaki, and T.~Yanagida, hep-ph/9707201, to appear
  in Phys.~Lett.~B; \\
  Izawa~K.-I., hep-ph/9708315; \\
  M.~Kawasaki, N.~Sugiyama, and T.~Yanagida, hep-ph/9710259.

\bibitem{Lyt}
  For a review, D.H.~Lyth, hep-ph/9609431.

\end{thebibliography}
\end{document}